\documentclass[a4paper,twocolumn,letter]{jpsj3}
\usepackage{color}
\usepackage{ulem}

\title{Photoemission Evidence for Valence Fluctuations and Kondo Resonance in YbAl$_2$}

\author{%
Masaharu~\textsc{Matsunami}$^{1, 2}$\thanks{E-mail: matunami@ims.ac.jp; Present address: UVSOR Facility, Institute for Molecular Science, Okazaki 444-8585, Japan}, 
Ashish~\textsc{Chainani}$^{2}$, 
Munetaka~\textsc{Taguchi}$^{2}$, 
Ritsuko~\textsc{Eguchi}$^{2}$\thanks{Present address: Graduate School of Natural Science and Technology, Okayama University, Okayama 700-8530, Japan}, 
Yasutaka~\textsc{Takata}$^{2}$, 
Masaki~\textsc{Oura}$^{2}$, 
Makina~\textsc{Yabashi}$^{2}$, 
Kenji~\textsc{Tamasaku}$^{2}$, 
Yoshinori~\textsc{Nishino}$^{2}$\thanks{Present address: Research Institute for Electronic Science, Hokkaido University, Sapporo 060-0812, Japan}, 
Tetsuya~\textsc{Ishikawa}$^{2}$, 
Masashi~\textsc{Kosaka}$^{3}$, and 
Shik~\textsc{Shin}$^{1, 2, 4}$
}
\inst{
$^{1}$Institute for Solid State Physics, University of Tokyo, Kashiwa, Chiba 277-8581, Japan \\
$^{2}$RIKEN SPring-8 Center, Sayo-cho, Sayo-gun, Hyogo 679-5148, Japan \\
$^{3}$Graduate School of Science and Engineering, Saitama University, Saitama 338-8570, Japan \\
$^{4}$Japan Science and Technology Agency (JST), Core Research for Evolutional Science and Technology (CREST), Chiyoda-ku, Tokyo 102-0075, Japan
}
\abst{
We use hard x-ray photoemission spectroscopy (HAXPES) to investigate the electronic structure of YbAl$_2$, for which the Yb valence has not been consistently reported to date. 
The bulk sensitivity and the analytical simplicity provided by the Yb\,3$d$ core-level HAXPES allow a reliable determination of the mean valence of Yb ions. 
For YbAl$_2$, it is evaluated to be +2.20, which remains nearly unchanged below 300\,K. 
The Kondo resonance peak with an extremely high Kondo temperature (above 2000\,K) is clearly identified in the valence-band spectra. 
The results indicate that a coherent Kondo state can be robust even in a nearly divalent system. 
}

\kword{YbAl$_2$, valence fluctuation, photoemission spectroscopy, HAXPES}

\begin{document}
\maketitle

For a long time, rare-earth intermetallic compounds have attracted much attention, since they exhibit a wide variety of strongly correlated electronic phenomena associated with the valence instability of these rare-earth ions. \cite{Varma,Lawrence} 
Recently, the concept of valence fluctuations has received renewed interest in relation to the unconventional quantum critical phenomena, particularly for Yb-based compounds such as YbRh$_2$Si$_2$ \cite{YbRh2Si2_Trovarelli} or $\beta$-YbAlB$_4$. \cite{YbAlB4_Nakatsuji} 
Theoretically, the role of the critical valence fluctuations has been invoked to explain their properties. \cite{Watanabe} 
The importance of critical valence fluctuations provides us with an opportunity to reexamine the valence fluctuation phenomena. 
Experimentally, the most fundamental issue is to precisely determine the mean valence, which is a measure of the $f$-electron occupation and localization.

Hard x-ray photoemission spectroscopy (HAXPES) with an excitation energy of 5--8\,keV is one of the best-suited techniques. 
Apart the well-known bulk sensitivity, HAXPES can probe the Yb\,3$d$ core levels, located at a binding energy ($E_{\rm B}$) of 1500--1600\,eV, which is usually not possible with laboratory sources. 
For the Yb\,3$d$ core-level spectra, the divalent (4$f^{14}$, fully occupied) and the trivalent (4$f^{13}$, one-hole localized) features are well separated, thus allowing highly reliable valence determination. 
Indeed, HAXPES has been applied to various Yb-based compounds, such as 
the valence transition system YbInCu$_4$, \cite{YbInCu4_Sato,RIXS_vs_HXPES,YbInCu4_Suga} 
the valence fluctuation systems YbAl$_3$ \cite{RIXS_vs_HXPES,YbAl3_HXPES_Suga} and YbCu$_2$Si$_2$, \cite{RIXS_vs_HXPES,Yb2+_Matsunami} 
the divalent systems Yb metal and YbS, \cite{Yb2+_Matsunami} 
the Kondo semiconductor YbB$_{12}$, \cite{YbAlB4_Yamaguchi} 
and the heavy-fermion superconductor YbAlB$_4$. \cite{YbAlB4_Okawa} 
In this letter, we demonstrate a precise determination of the Yb valence for YbAl$_2$, which has also been considered as a prototypical valence fluctuation system.

YbAl$_2$ crystallizes in the cubic Laves MgCu$_2$ structure. 
The electrical resistivity shows a normal metallic behavior without a coherence peak in the temperature ($T$) range below 300\,K. \cite{Havinga,Tsujii,Gorlach,Nowatari} 
The electronic specific-heat coefficient, which provides a measure of effective electron mass, exhibits a slightly enhanced value, $\sim$10--17\,mJ/K$^2$mol \cite{Tsujii,Gorlach,Nowatari}. 
The magnetic susceptibility has a broad maximum ($T_{\rm max}$) at 800--850\,K and is nearly independent of $T$ below 300\,K after subtracting the impurity contribution. \cite{Latt1_Iandelli,Havinga,Gorlach,Klaasse} 
The Kondo $T$ ($T_{\rm K}$) has been estimated as 2000--2600\,K from the 3$T_{\rm max}$ or the inelastic neutron scattering. \cite{Neutron_Murani} 
The Yb valence has been investigated by some techniques such as the lattice parameters \cite{Latt1_Iandelli,XAS1Latt2_Bauchspeiss} and the x-ray absorption spectroscopy (XAS) at the Yb-$L_{\rm III}$ edge. \cite{XAS1Latt2_Bauchspeiss,XAS2_Rohler,XAS3_Dallera} 
In addition, the valence-band photoemission spectroscopy (PES) using $h\nu$=\,70 \cite{PES1_Kaindl}, 1487 (Al-$K\alpha$) \cite{PES2_Oh}, and 1254\,eV (Mg-$K\alpha$) \cite{PES3_Abbati} has also been applied to study Yb valence. 
The reported valence values, however, have considerably large variation between +2.0 and +2.6. 
Such a large discrepancy may be caused by surface effects \cite{Probing_Depth}, surface oxidation, or inconsistent precision of analysis. 
After carefully overcoming such considerations, our HAXPES results provide a conclusive Yb mean valence and consistency with its estimated $T_{\rm K}$. 
Furthermore, the Kondo resonance peak is identified in the valence-band spectra, indicating that a coherent Kondo ground state can be robust even in a nearly divalent system such as YbAl$_2$.

The single crystals of YbAl$_2$ were grown by the lithium flux method \cite{Nowatari}. 
Clean sample surfaces were obtained by fracturing $in$ $situ$. 
Synchrotron-based HAXPES experiments were carried out at the undulator beamline BL29XUL in SPring-8, using $h\nu$=7940\,eV and a hemispherical electron analyzer, Scienta R4000-10\,kV. \cite{BL29XUL} 
The total energy resolution was set to 250 and 170\,meV for the measurements of the core-level and valence-band spectra, respectively. 
To obtain the higher-resolution (50\,meV) valence-band spectrum, we also performed soft x-ray (SX-) PES at the undulator beamline BL17SU using $h\nu$=580\,eV and Scienta SES-2002. \cite{BL17PES} 
The $E_{\rm B}$ of samples was calibrated using the Fermi level ($E_{\rm F}$) of an evaporated Au film.

\begin{figure}[t]
\begin{center}
\includegraphics[width=0.48\textwidth,clip]{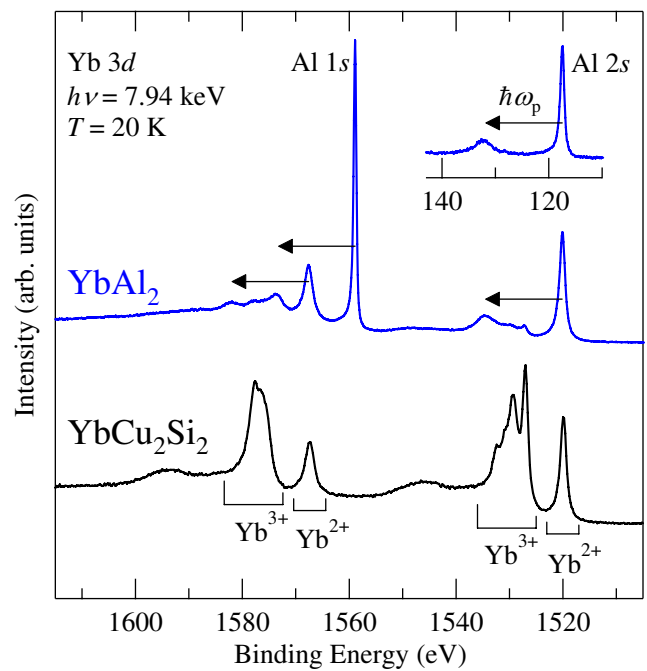}
\end{center}
\caption{
(Color online) 
Yb\,3$d$ core-level HAXPES spectrum including the Al\,1$s$ core level for YbAl$_2$ in comparison with that for YbCu$_2$Si$_2$ \cite{Yb2+_Matsunami}. 
Inset shows the Al\,2$s$ core-level spectrum, which is plotted with the same energy scale as the main panel, with the satellite due to plasmon excitations. 
Arrows represent the plasmon energy $\hbar\omega_{\rm p}$=14.32\,eV. 
}
\end{figure}

Figure 1 shows the Yb\,3$d$ core-level spectrum of YbAl$_2$ at 20\,K, compared with that of YbCu$_2$Si$_2$, which is known to show a Yb valence of +2.8. \cite{Yb2+_Matsunami} 
The sharp peak at $E_{\rm B}$=1559\,eV is due to the Al\,1$s$ core level. 
The Yb\,3$d$ core-level spectra are separated into the 3$d_{5/2}$ region at 1515--1540\,eV and the 3$d_{3/2}$ region at 1560--1585\,eV owing to spin-orbit splitting. 
Both the spin-orbit components in YbCu$_2$Si$_2$ show two final-state configurations: the 3$d^9$4$f^{14}$ (Yb$^{2+}$) line at $\sim$1520 and $\sim$1568\,eV, and the 3$d^9$4$f^{13}$ (Yb$^{3+}$) multiplets at $\sim$1525--1538 and $\sim$1572--1583\,eV, which are due to the valence fluctuations. 
Similarly for YbAl$_2$, a clear Yb$^{2+}$ line and weak Yb$^{3+}$ multiplets can be observed, although we also find the Al\,1$s$ core level positioned between 3$d_{5/2}$ and 3$d_{3/2}$ regions and overlapped satellite features. 
The satellite features are caused by the energy loss due to plasmon excitations ($\hbar\omega_{\rm p}$=14.32\,eV), as confirmed by measuring the Al\,2$s$ core-level spectrum shown in the inset of Fig.~1.

\begin{figure}[t]
\begin{center}
\includegraphics[width=0.48\textwidth,clip]{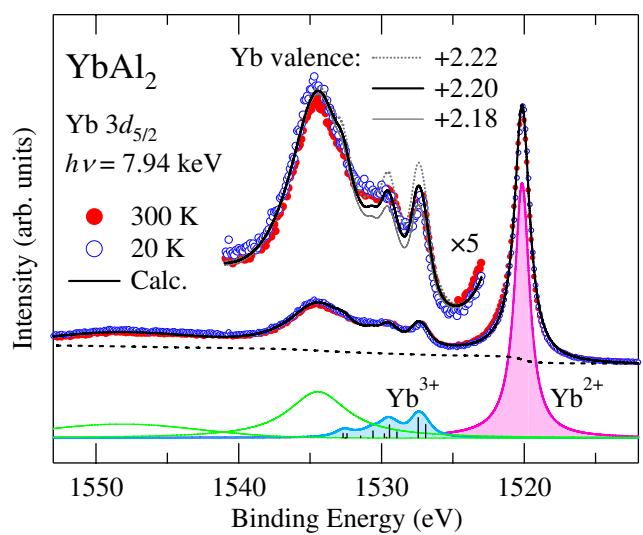}
\end{center}
\caption{
(Color online) 
Comparison of the experimental spectra with the calculation by the atomic-multiplet model for the Yb\,3$d_{5/2}$ core level including the energy-loss satellites.  
The atomic multiplets are broadened in both terms of lifetime and experimental resolution. 
Dashed curve represents the integral background. 
In the expanded panel, the calculated reference spectra with Yb valence of +2.20$\pm$0.02 are also plotted. 
}
\end{figure}

In order to determine the Yb mean valence for YbAl$_2$, the obtained spectra were analyzed in terms of the atomic multiplet model for free Yb$^{2+}$ and Yb$^{3+}$ ions. \cite{Atomic} 
Figure 2 shows the Yb\,3$d_{5/2}$ core-level spectra at 300 and 20\,K, in comparison with those obtained from the atomic-multiplet calculation. 
The calculated discrete spectra were broadened by a Doniach-{\u{S}}unji{\'c} function convoluted with a Gaussian representing the experimental energy resolution. 
The energy-loss satellites were fitted with a Lorentzian for the relatively intense peak at $\sim$1534\,eV and a Gaussian for the broad peak at $\sim$1548\,eV. 
As a background, the standard integral type one was used. 
The experimental spectra for both $T$ are well reproduced by the calculation with the Yb mean valence of +2.20, which is obtained from the weight ratio between Yb$^{2+}$ and Yb$^{3+}$ components. 
As can be confirmed in the expanded panel in Fig. 2, the experimental spectra are located well between the calculated reference spectra, and hence, the accuracy in determining the Yb mean valence is estimated as $\sim\pm$0.02.

\begin{figure}[t]
\begin{center}
\includegraphics[width=0.48\textwidth,clip]{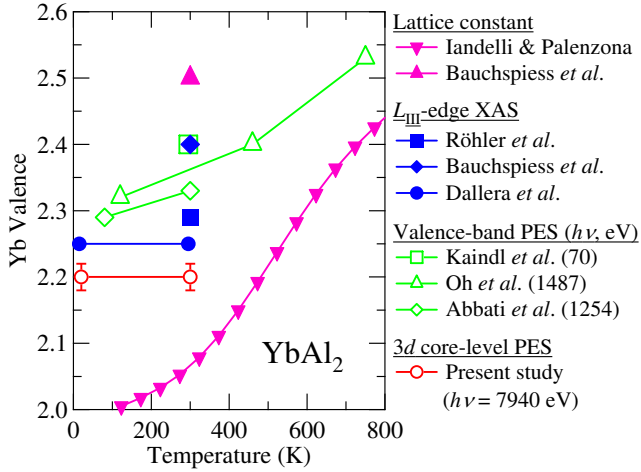}
\end{center}
\caption{
(Color online) 
$T$ dependence of Yb valence in YbAl$_2$ evaluated from the lattice parameters \cite{Latt1_Iandelli,XAS1Latt2_Bauchspeiss},the XAS at $L_{\rm III}$ edge \cite{XAS1Latt2_Bauchspeiss,XAS2_Rohler,XAS3_Dallera}, the valence-band PES \cite{PES1_Kaindl,PES2_Oh,PES3_Abbati}, and the 3$d$ core-level HAXPES in the present study. 
}
\end{figure}

Figure 3 shows the $T$-dependent Yb valence in YbAl$_2$ evaluated from the lattice parameters \cite{Latt1_Iandelli,XAS1Latt2_Bauchspeiss},the XAS at the $L_{\rm III}$ edge \cite{XAS1Latt2_Bauchspeiss,XAS2_Rohler,XAS3_Dallera}, the valence-band PES, \cite{PES1_Kaindl,PES2_Oh,PES3_Abbati} and the above-described HAXPES. 
In the following, we discuss the origin of the large discrepancy among the results of these techniques. 
Since the lattice parameters as well as the magnetic susceptibility probe an average signal of Yb$^{2+}$ and Yb$^{3+}$ components, it is intrinsically difficult to estimate a precise Yb valence. 
This difficulty may be related to the large variation among two results for YbAl$_2$. \cite{Latt1_Iandelli,XAS1Latt2_Bauchspeiss} 
On the other hand, using spectroscopies such as XAS and PES, each component can be separately probed. 
The Yb $L_{\rm III}$-edge XAS is a well-known powerful tool for investigating the electronic structure, including the Yb-valence determination, and can be performed even under magnetic field and/or external pressure. 
However, Yb$^{2+}$ and Yb$^{3+}$ components in spectra are relatively broad and overlap significantly with each other, thus limiting the reliability of Yb valence determination compared with the Yb\,3$d$ core-level PES. 
Moreover, in the case of powder samples, as are used in XAS studies under high pressure, \cite{XAS1Latt2_Bauchspeiss,XAS3_Dallera} surface oxidation effects can have a serious impact on the Yb valence even with the bulk sensitive probe. 
Such a problem has been pointed out in terms of an extra trivalent signal derived from the surface Yb$_2$O$_3$ layer in the HAXPES of a purely divalent system, YbS. \cite{Yb2+_Matsunami} 
For YbAl$_2$, the sharp rise at low $T$ in the magnetic susceptibility has also been attributed to the surface Yb$_2$O$_3$ as an impurity. \cite{Havinga,Gorlach,Klaasse} 
In our HAXPES measurements using $in$-$situ$-fractured single crystals of YbAl$_2$, no O-1$s$ signal has been detected. 
Therefore, the major difference in estimating Yb valence using XAS and our HAXPES results can be due to the surface oxidation of powder samples in XAS. 
Next, the HAXPES results are compared with previous lower photon energy valence-band PES studies. \cite{PES1_Kaindl,PES2_Oh,PES3_Abbati}
Since the valence-band spectra must include signals from orbitals other than Yb\,4$f$ electrons spread over a wide energy, it is difficult to precisely estimate the Yb valence, compared with the core-level spectroscopies. 
In the case of the low-energy PES, more importantly, the surface Yb$^{2+}$-4$f$ components can be observed, thus making it very difficult to obtain the precise bulk valence. 
For the previous PES with $h\nu$=70\,eV, \cite{PES1_Kaindl} such components might be overestimated, leading to a larger value of bulk valence than the intrinsic value. 
Subsequent SX-PES studies with $h\nu$=1487 \cite{PES2_Oh} and 1254\,eV \cite{PES3_Abbati} also assumed extra surface Yb$^{2+}$-4$f$ components, which could not be separated with the bulk ones because of the limited resolution, and hence also estimated a relatively large bulk valence. 
Actually, even in SX-PES with lower-energy $h\nu$=580\,eV for YbAl$_2$, the surface Yb$^{2+}$-4$f$ components are negligibly small, as shown later. 
As a result, the simplicity of the experimental data and its analysis for bulk sensitive Yb\,3$d$ core-level HAXPES provides a high reliability in Yb valence determination.

The $T$ independence of the Yb valence below 300\,K is a distinct behavior compared with the other valence fluctuation systems such as YbAl$_3$ \cite{RIXS_vs_HXPES,YbAl3_HXPES_Suga} and YbCu$_2$Si$_2$. \cite{RIXS_vs_HXPES} 
It can be understood in terms of the extremely high $T_{\rm K}$ (=2000--2600\,K) in YbAl$_2$, since the valence should vary significantly around $T_{\rm K}$. 
This finding is consistent with the electrical resistivity showing no coherence peak \cite{Havinga,Tsujii,Gorlach,Nowatari} and the constant magnetic susceptibility below 300\,K. \cite{Latt1_Iandelli,Havinga,Gorlach}

\begin{figure}[t]
\begin{center}
\includegraphics[width=0.48\textwidth,clip]{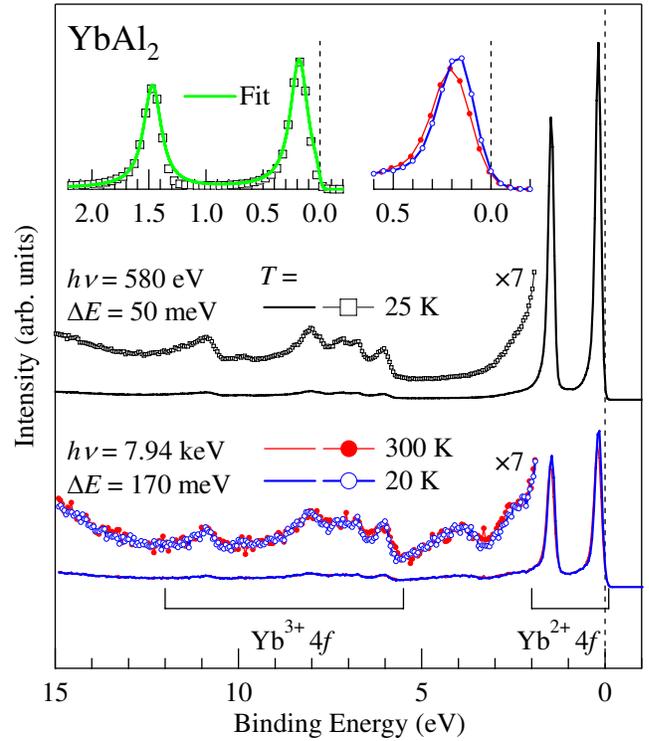}
\end{center}
\caption{
(Color online) 
$T$ dependence of valence-band HAXPES spectra and the high-resolution SX-PES spectrum at 25\,K of YbAl$_2$. 
Left inset shows the fitting result of the Yb$^{2+}$-4$f$ doublet in the SX-PES spectrum. 
Right inset shows the expanded scale of the HAXPES spectra in the near-$E_{\rm F}$ region. 
}
\end{figure}

We now discuss the electronic structure of YbAl$_2$ on the basis of the valence-band spectra, which also reflects the fluctuating Yb valence. \cite{Valence_VBPES} 
Figure 4 shows the $T$-dependent valence-band HAXPES spectra and the high-resolution SX-PES spectrum at 25\,K, which are normalized for integrated intensity up to $E_{\rm B}$=13\,eV. 
The most prominent features are two sharp peaks between $E_{\rm F}$--2\,eV, which are derived from the Yb$^{2+}$-4$f$ doublet ($J$=7/2 and 5/2) corresponding to the 4$f^{13}$ final state. 
Next, two weak structures are observed at 2--5\,eV in the HAXPES spectra, but not in the SX-PES spectrum. 
According to the band calculation \cite{Optical_Lee} and taking into account the photoionization cross sections \cite{CrossSec}, these can be attributed to Al\,3$s$ and 3$p$ states. 
The multiple peak feature observed at 5--12\,eV is due to the Yb$^{3+}$-4$f$ multiplet corresponding to the 4$f^{12}$ final state. 
The existence of a fractional Yb$^{3+}$ component is consistent with the fact that YbAl$_2$ is not a purely divalent system. 
It is important to note that the Yb$^{2+}$-4$f_{7/2}$ peak crosses $E_{\rm F}$, in contrast to the case of purely divalent Yb systems in which this peak is located significantly below $E_{\rm F}$. \cite{Yb2+_Matsunami} 
In order to analyze this peak, the Yb$^{2+}$-4$f$ doublet in the SX-PES spectrum was fitted by two Doniach-{\u{S}}unji{\'c} functions multiplied with a Fermi-Dirac distribution at $T$=25\,K and convoluted with a Gaussian, as shown in the left inset of Fig.~4. 
The spectrum is reproduced well by this simple fitting, which does not include any background and/or surface components. 
The obtained position and the intrinsic width (FWHM) of the 4$f_{7/2}$ peak are 180 and 170\,meV, respectively, both of which are in excellent agreement with the estimated $T_{\rm K}$ (=2000--2600\,K) and, in particular, the characteristic peak position (=180\,meV) in inelastic neutron scattering. \cite{Neutron_Murani} 
Therefore, the Yb$^{2+}$-4$f_{7/2}$ peak can be identified as the Kondo resonance peak, thereby providing conclusive evidence for Kondo screening in YbAl$_2$ and consistency with the $T_{\rm max}$ behavior in the magnetic susceptibility. \cite{Havinga,Klaasse} 
This finding establishes that the coherent Kondo ground state can be robust even in a nearly divalent regime.

Although the $T$ dependence is small in the valence-band spectra as well as the core-level spectra, the Yb$^{2+}$-4$f$ peaks slightly shift by $\sim$30\,meV toward lower $E_{\rm B}$ with cooling, as shown in the right inset of Fig.~4, in contrast to the unshifted behavior of the Yb$^{3+}$-4$f$ multiplet. 
Such $T$ variation has also been observed in YbAl$_3$, and it was suggested to be a signature of the deviation from the framework of the single-impurity Anderson model. \cite{YbAl3_HXPES_Suga} 
In addition, the weight of the Yb$^{2+}$ component slightly increases with cooling, suggesting a slight reduction of Yb valence within the error bar. \cite{Valence_VBPES} 
It is noted that since the Kondo effect is intrinsically associated with the hybridization between conduction band and $f$-electron states and since the estimated $T_{\rm K}$ of YbAl$_2$ is very high, it is expectedly difficult to estimate the small change in $T$-dependent hybridization between 20 and 300\,K.

In conclusion, we performed HAXPES and SX-PES on YbAl$_2$. 
From the Yb\,3$d$ core-level spectra, the mean valence of Yb ions was evaluated to be +2.20, which remained nearly unchanged below 300\,K. 
Through the comparison with those obtained by other techniques, the reliability of the valence determined by HAXPES was demonstrated. 
Furthermore, the Kondo resonance peak was identified in the valence-band spectra, indicating that a coherent Kondo ground state can be robust even in a nearly divalent system such as YbAl$_2$.

\begin{acknowledgment}

We thank H. Okamura and S. Kimura for valuable comments. 
The measurements of HAXPES and SX-PES were carried out with the approval of the RIKEN SPring-8 Center (Proposals No. 20090041 and No. 20090033, respectively). 
This research is supported by the Japan Society for the Promotion of Science (JSPS) through its Funding Program for World-Leading Innovative R\&D on Science and Technology (FIRST) Program. 

\end{acknowledgment}


\begin{thebibliography}{99}
\bibitem{Probing_Depth}
The probing depth in low-energy (VUV-SX) PES, HAXPES, and $L_{\rm III}$-edge (fluorescence-yield) XAS is generally considered as $<$20 and $\sim$80\,\AA, and $\sim$10\,$\mu$m, respectively.
\bibitem{Valence_VBPES}
The Yb mean valence estimated from the valence-band spectra in Fig.~4 is +2.30$\pm$0.05 (20\,K) and +2.31$\pm$0.05 (300\,K) for HAXPES and +2.13$\pm$0.03 (25\,K) for SX-PES. 
The discrepancy with the valence estimated with Yb\,3$d$ core-level HAXPES can be attributed to the existence of orbitals other than Yb\,4$f$ electrons spread over a wide energy, which cannot be accurately separated, in valence-band spectra.
\end{thebibliography}
\end{document}